\documentclass{sf2a-conf2013}
\usepackage{graphicx}
\usepackage{hyperref}
\usepackage[]{natbib}
\usepackage{epstopdf}

\def\BibTeX{{\rm B\kern-.05em{\sc i\kern-.025em b}\kern-.08em
    T\kern-.1667em\lower.7ex\hbox{E}\kern-.125emX}}
\bibpunct{(}{)}{;}{a}{}{,}  

\newcommand{\ind}[1]{_{\mathrm{#1}}}
\newcommand{\diff}{\mathrm{d}}

\def\Kepler{\emph{Kepler}}

\def\numax{\nu\ind{max}}\def\nmax{n\ind{max}}

\def\dnuenv{\delta\nu\ind{env}}
\def\nenv{n\ind{env}}
\def\d0l{d_{0\ell}}
\def\Dnu{\Delta\nu}

\def\Dnuobs{\Delta\nu\ind{obs}}
\def\Dnuas{\Delta\nu\ind{as}}
\def\Dnuasobs{\Delta\nu\ind{as,obs}}
\def\Dnuasmod{\Delta\nu\ind{as,mod}}
\def\numaxobs{\nu\ind{max,obs}}
\def\numaxmod{\nu\ind{max,mod}}
\def\Dnurho{\nu_0}
\def\Tg{\Delta\Pi_1}

\def\epsobs{\varepsilon\ind{obs}}

\def\Dnuas{\Delta\nu\ind{as}}
\def\epsas{\varepsilon\ind{as}}

\def\Hmax{H\ind{max}}
\def\Bmax{B\ind{max}}
\def\Amax{A\ind{max}}

\def\Robs{R\ind{obs}} 
\def\Mobs{M\ind{obs}}
\def\Ras{R\ind{as}}
\def\Mas{M\ind{as}}

\def\ng{n\ind{g}}

\newcommand\iref{_\odot}

\newcommand\dnusas{\Dnu\ind{ref}}
\newcommand\numaxas{\nu\ind{ref}}
\newcommand\Teff{{T\ind{eff}}}
\newcommand\Ts{T\iref}
\newcommand\Rs{R\iref}
\newcommand\Ms{M\iref}
\newcommand\Rsis{R} 
\newcommand\Msis{M} 

\renewcommand{\baselinestretch}{.97}

\begin{document}

\TitreGlobal{SF2A 2013}

\title{Red giants seismology}
\runningtitle{Red giants seismology}
\author{B. Mosser, R. Samadi, K. Belkacem}\address{LESIA, CNRS, Universit\'e Pierre et Marie Curie,
Universit\'e Denis Diderot, Observatoire de Paris, 92195 Meudon
cedex, France; \email{benoit.mosser@obspm.fr}}

\setcounter{page}{1}

\maketitle

\begin{abstract}
The space-borne missions CoRoT and \emph{Kepler} are indiscreet.
With their asteroseismic programs, they tell us what is hidden
deep inside the stars. Waves excited just below the stellar
surface travel throughout the stellar interior and unveil many
secrets: how old is the star, how big, how massive, how fast (or
slow) its core is dancing. This paper intends to \emph{paparazze}
the red giants according to the seismic pictures we have from
their interiors.
\end{abstract}

\begin{keywords}
Stars: oscillations -- Stars: interiors -- Stars: evolution --
Methods: data analysis
\end{keywords}


\section{Introduction}

\begin{figure}[ht!]
 \centering
 \includegraphics[width=0.8\textwidth,clip]{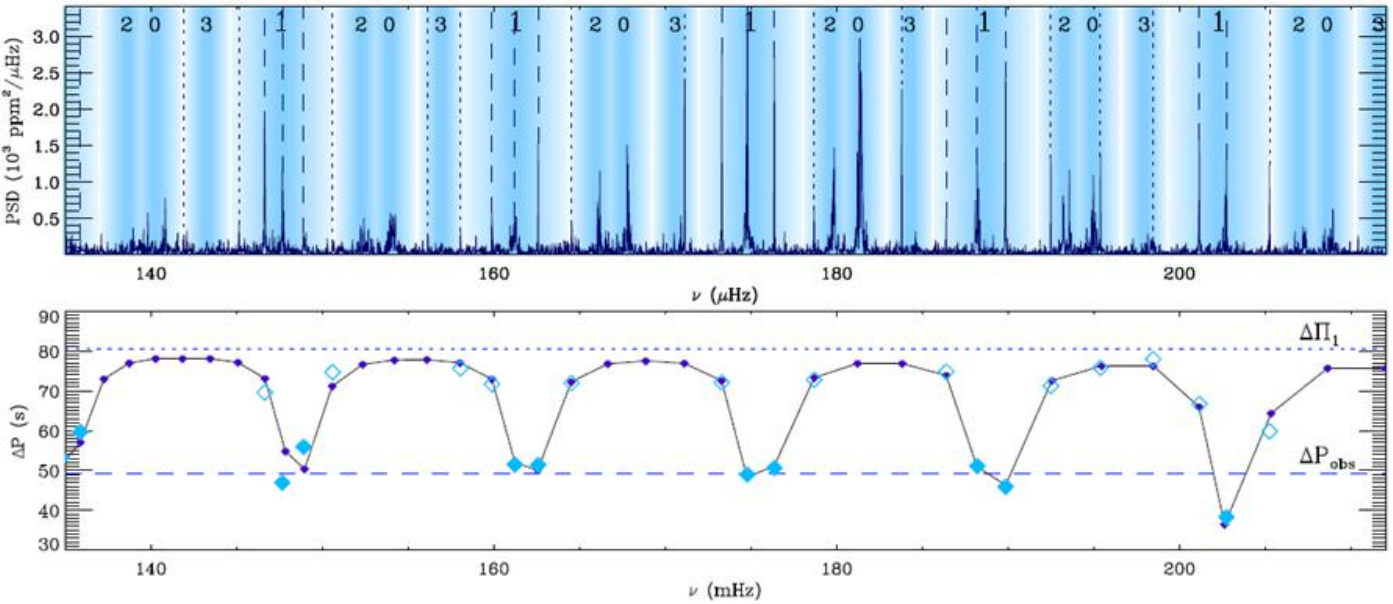}
  \caption{\textbf{Top:} Typical red giant oscillation pattern. Radial ($\ell=0$) and quadrupole ($\ell=2$) follow a regular comb pattern.
  \textbf{Bottom:} Period spacing between dipole modes; they have a mixed character and therefore show an irregular pattern;
  they can be identified with the asymptotic expansion. Figure from \cite{2012A&A...540A.143M}.}
  \label{example_mixte}
\end{figure}

As denoted by many authors, red giant seismology is an exquisite
surprise provided by the space missions CoRoT and \Kepler. The
analysis of a wealth of light curves recorded with unique length,
continuity and photometric precision has already revealed many
secrets. The most striking results, up to now, are provided by the
observation of mixed modes (Fig.~\ref{example_mixte}). Such modes
result from the coupling of gravity waves, propagating in the
radiative core region, with pressure waves propagating in the
stellar envelope. They directly reveal information from the
stellar core: the nature of the nuclear reaction
\citep{2011Natur.471..608B} and the mean core rotation rate
\citep{2012Natur.481...55B}.

An analysis of red giant seismic observational results has been
given in \cite{2013EPJWC..4303003M}. Tricky points related to the
data analysis are presented in a companion paper \citep{roscoff},
where emphasis is given on red giant interior structure
as revealed by asteroseismology. Only ensemble asteroseismology
results are presented in this paper, obtained from the monitoring
of a cohort of stars. Analysis and modelling of individual stars
are not considered. They have started for a handful of targets
\citep{2011MNRAS.415.3783D,2011ApJ...742..120J,2012A&A...538A..73B}.
Such individual analysis are crucial for enhancing the
understanding of the stellar interior structure and of the
physical input to be considered, such as the measurement of the location of the helium
second-ionization region \citep{2010A&A...520L...6M}, or the
measurement of differential rotation
\citep{2012Natur.481...55B,2012ApJ...756...19D}.

In this work, ensemble asteroseismology results are derived from
the observations of evolutionary sequences (Section
\ref{sequence}). For red giants, asteroseismology benefits from
the large homology of their interior structure, which translates
into homologous properties of the oscillation spectrum (Section
\ref{homology}). In this Section, we also show how mixed-modes
directly probe the stellar cores, and then investigate how stellar
evolution is monitored by seismology on the RGB and in the red
clump. Scaling relations derived from the homologous properties of
red giants are discussed in Section \ref{scaling}, with a special
emphasis on the mass and radius scaling relations and on their
calibration. Rotation is discussed in Section \ref{rotation}. A
recent leap toward oscillations detected in semi-regular variables
is presented in Section \ref{ogle}.

\section{Evolutionary sequences\label{sequence}}

\begin{figure}[ht!]
 \centering
\includegraphics[width=0.9\textwidth,clip]{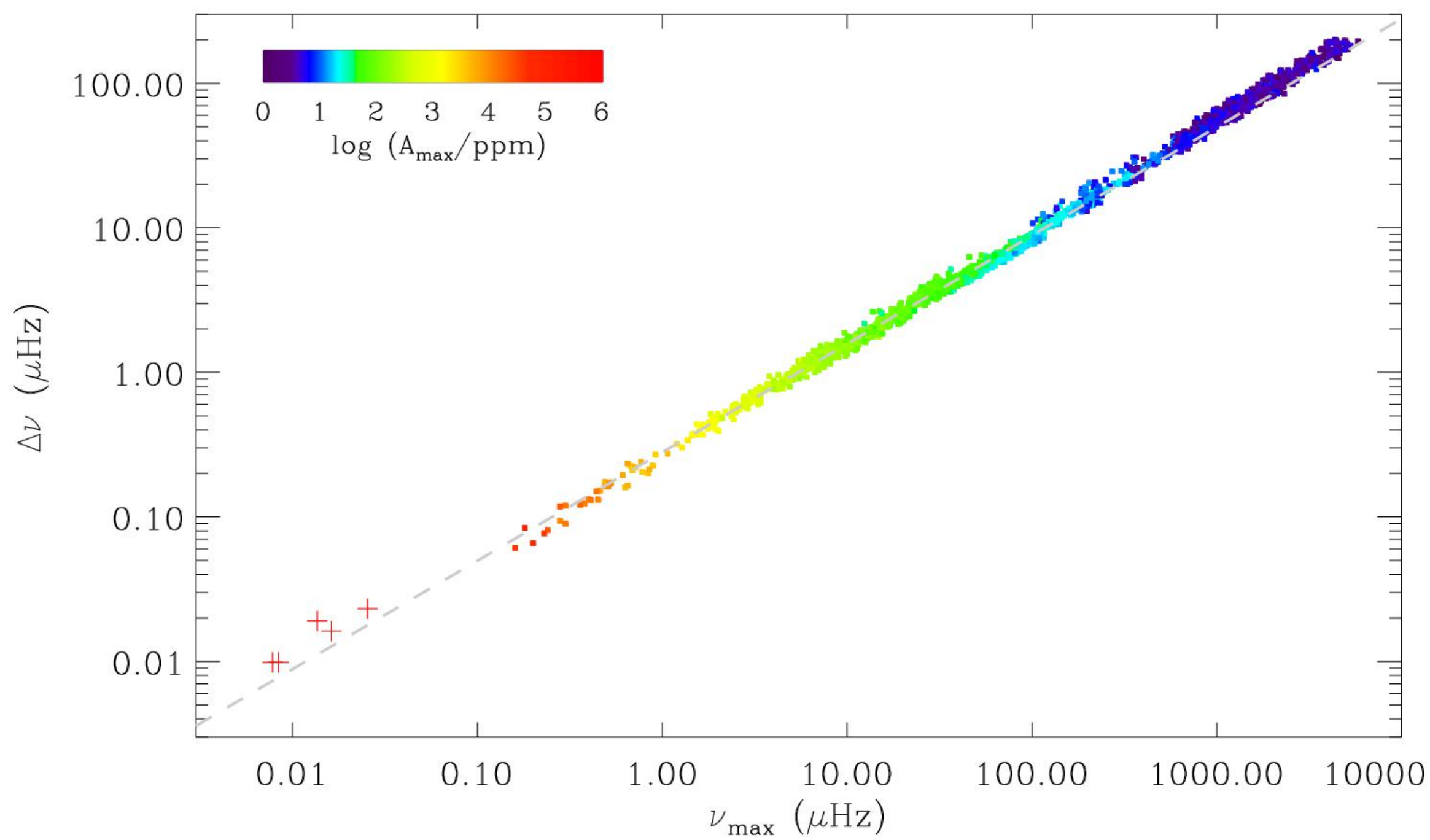}
  \caption{$\numax$ -- $\Dnuobs$ relation, with data from \cite{2011MNRAS.415.3539V}, \cite{2012A&A...540A.143M}, and \cite{mosser_ogle}.
  Plusses are red supergiants showing at least two oscillation modes observed by
  \cite{2006MNRAS.372.1721K}, with $\Dnuobs$ and $\numax$
  derived from a combination of the frequencies.
  The colors code the mean maximum amplitude $\Amax$ of the radial oscillations. The grey dashed line has a slope 3/4, in
  agreement with Eq.~\ref{scaling-dnu}}
  \label{numaxdnu}
\end{figure}

The observation of thousands of red giants with CoRoT
\citep[e.g.,][]{2009A&A...506...57D,2010A&A...517A..22M} and
\Kepler\ \citep[e.g.,][]{2010ApJ...723.1607H,2013ApJ...765L..41S}
allows us to address ensemble asteroseismology. According to the
distance measurements derived from the scaling relations, red
giants with magnitudes up to 13 are monitored vast regions of the
Galaxy (Section \ref{distancepop}). They show a large variety of
metallicity \citep{2012MNRAS.423..122B}. Independent of a specific
population analysis \citep[e.g.,][for a one-solar-mass
evolutionary sequence]{2011ApJ...740L...2S}, we note that,
apparently, the cohort of stars provided by both missions provide
evolution sequence  of low-mass stars. A reason justifying that we
address sequences of evolution is provided by the global
asteroseismic parameters $\numax$ and $\Dnuobs$: $\numax$ is the
frequency of maximum oscillation signal; $\Dnuobs$ is the mean
large frequency separation of radial modes observed around
$\numax$ \citep[e.g.,][]{2009A&A...508..877M}. The frequency
$\numax$ provides a proxy of the acoustic cutoff frequency, hence
of $M\ R^{-2}\, \Teff^{-1/2}$ \citep{2011A&A...530A.142B}.
Following \cite{2013arXiv1307.3132B}, we consider that $\Dnuobs$
provides an acceptable proxy of the dynamical frequency $\Dnurho$
that scales as $\sqrt{\mathcal{G}M/R^3}$. From these scalings, we
derive
\begin{equation}\label{scaling-dnu}
   \Dnuobs \simeq  \Dnurho  \propto M^{-1/4} \ \Teff^{3/8} \
   \numax^{3/4} .
\end{equation}
On the main sequence, the frequency $\Dnuobs$ scales as
$\numax^{0.8}$ \citep{2011MNRAS.415.3539V}. The discrepancy
between this exponent and the 3/4 value in Eq.~\ref{scaling-dnu}
is due to the fact that low-mass and high-mass evolution tracks
are in different regions of the main sequence. On the contrary,
the observed scaling exponent on the RGB is much closer to 3/4
(Fig.~\ref{numaxdnu}) since the RGB regroups the evolution of
low-mass stars. The agreement between Eq.~\ref{scaling-dnu} and
global oscillation parameters observed over more than four decades
in the red giant regime, completed by measurements on red
supergiant stars \citep{2006MNRAS.372.1721K}, indicates that the
stellar red giant populations observed by CoRoT or \Kepler\
constitute a set of stars homogenous enough to mimic stellar
evolution, then justifying the relevance of the scaling relations.
The difference between the observed exponent and 3/4 is small
enough to be interpreted either by the very low variation of
$\Teff$ with $\numax$, or by the difference between $\Dnuobs$ and
$\nu_0$, and by  other parameters that need to be calibrated but
necessary have a limited influence. Masses in the range $0.9$ --
$2\,M_\odot$ are present at all stages, hence for all $\numax$, so
that the mass parameter in Eq.~\ref{scaling-dnu} plays no
significant role.

Before addressing the scaling relations, we discuss about stellar
homology, seen by both radial and non-radial oscillation modes.

\section{Homology\label{homology}}

\subsection{From interior structure homology to seismic homology}

\begin{figure}[t!]
 \centering
 \includegraphics[width=0.8\textwidth,clip]{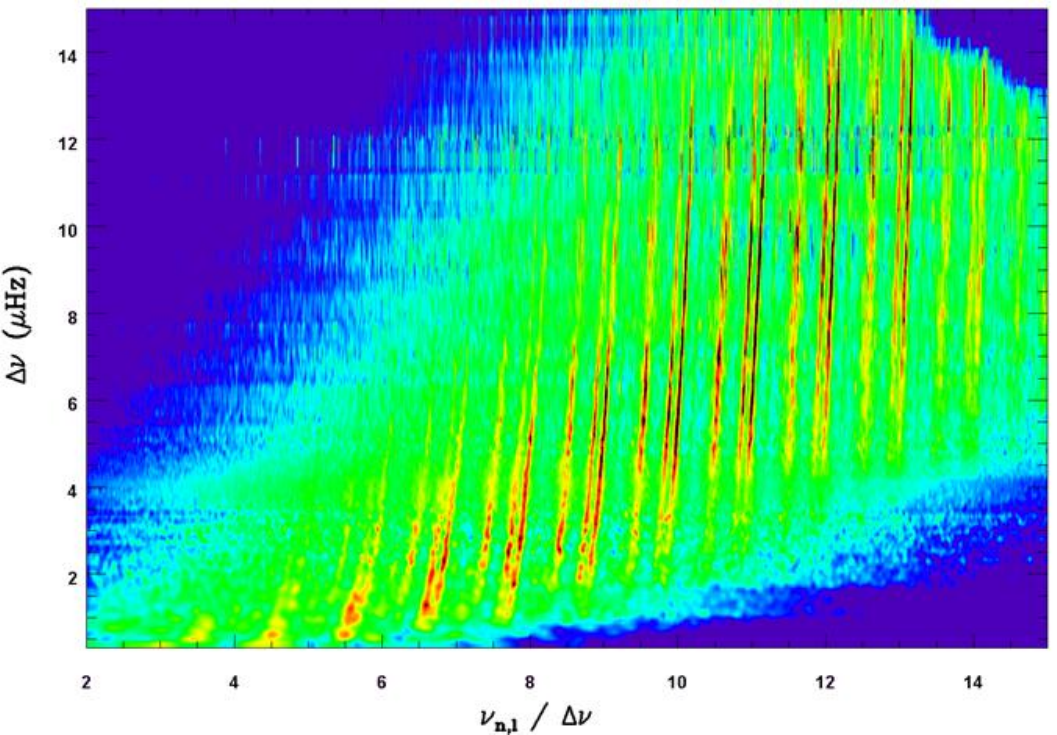}
  \caption{\Kepler\ red giant oscillation spectra, as a function
  of the dimensionless frequency $\nu_{n,\ell}/\Dnu$,
  sorted by increasing large separation (y-axis). The colors code the
  oscillation amplitude normalized to the maximum oscillation signal.
  The close parallel ridges are drawn by quadrupole and radial modes,
  from the radial orders 3 to 13; the width of the dipole ridges is wider, due
  to the presence of mixed modes. }
  \label{mosser_universel}
\end{figure}

Red giant interior structure is divided in three main regions; the
dense helium core,  the surrounding  thin hydrogen-burning shell,
and  the thick, mostly convective envelope
\citep{1990sse..book.....K}. Homology is ensured by the fact that
this structure is dominated by generic physics: the
thermodynamical conditions of the hydrogen-burning shell being
related on the one side with the helium core and on the other side
with the convective envelope, the core and envelope properties are
closely linked together \citep{2013ApJ...766..118M}. Consequently,
CoRoT observations have shown evidence of a very simple and useful
property of the red giant oscillation pattern: following the
interior structure homology, the oscillation pattern can also be
defined as homologous. The concept of \emph{universal red giant
oscillation pattern} was therefore introduced by
\cite{2011A&A...525L...9M}, as an alternative form to the usual
asymptotic expansion \citep{1980ApJS...43..469T}, with the
observed large separation as the only free parameter. The
second-order asymptotic expansion expresses, for low angular
degrees ($\ell\ll n$), as
\begin{equation}\label{ordre_deux}
    \nu_{n,\ell} = \left( n+\epsobs (\Dnuobs) + \d0l (\Dnuobs) + {\alpha \over 2}\, (n-\nmax)^2\right) \
    \Dnuobs ,
\end{equation}
where the dimensionless parameter $\nmax$ is defined by
$\numax/\Dnuobs - \epsobs$. Homology is expressed by the
dependence in the observed large separation $\Dnuobs$ of the
offsets $\epsobs$ and $\d0l$. The radial offset $\epsobs$ helps
locate the radial ridge; the non-radial offsets $\d0l$ express the
shifts of the different degrees $\ell$ compared to the radial
modes \citep[e.g.,][]{2012ApJ...757..190C}. Homology is
illustrated in Fig.~\ref{mosser_universel}, where \Kepler\ red
giant oscillation spectra are plotted on the same graph, with a
dimensionless frequency in abscissa, and sorted by increasing
large separation values. The alignment of the ridges, each one
corresponding to a given radial order $n$ and angular degree
$\ell$, demonstrates the universality of the oscillation pattern.

The form of Eq.~\ref{ordre_deux}, which includes a
quadratic term, accounts for the measurement of $\Dnuobs$ around
$\numax$ in non-asymptotic conditions. The asymptotic form writes
\begin{equation}\label{ordre_deux_as}
    \nu_{n,\ell} = \left( n'_\ell + {A_\ell \over n'_\ell}\right) \
    \Dnuas \hbox{ \ with \ } n'_\ell = n + \epsas + {\ell\over 2}
\end{equation}
The asymptotic large separation $\Dnuas$ is slightly greater than
the observed value. The link between the observed and asymptotic
parameters is explored in \cite{2013A&A...550A.126M}. The
asymptotic counterpart of $\epsobs$ is, for low-mass stars,
$\epsas \equiv 1/4$ \citep{1980ApJS...43..469T}. The high accuracy
level reached by Eq.~\ref{ordre_deux} has been show in previous
comparison work
\citep{2011MNRAS.415.3539V,2011A&A...525A.131H,2012A&A...544A..90H}.
A quantitative analysis of the accuracy of the measurement of
$\Dnuobs$ with Eq.~\ref{ordre_deux} is done in \cite{roscoff}: the
uncertainty is about $0.02\,\mu$Hz for all evolutionary stages.

\subsection{Sounding the core\label{mixtes}}

Asteroseismology aims at probing the whole stellar interior
structure. In the Sun, such an analysis performs very efficiently,
but less efficiently in the core, since pressure modes do not
probe the core as efficiently as they probe the other regions, the
number of pressure modes sounding the solar core being low. This
explains the quest for solar gravity modes that precisely probe
the core \citep{2010A&ARv..18..197A}.

In red giants, g modes are not present, but appear indirectly
through the coupling of gravity waves propagating in the core with
pressure waves propagating in the envelope. A toy model for
explaining this coupling is given in \cite{roscoff}. Since the
first observation of mixed modes in red giants
\citep{2011Sci...332..205B}, a wealth of information has been
provided by their analysis. \cite{2011Natur.471..608B} and
\cite{2011A&A...532A..86M} have shown that different mixed-mode
patterns help distinguish helium-core burning giants in the red
clump from hydrogen-shell burning giants on the RGB.
\cite{2012A&A...540A.143M} have proposed that an asymptotic
expansion is able to depict the mixed-mode pattern. This
expansion, derived from the formalism developed by
\cite{1989nos..book.....U}, introduces the gravity period spacing
$\Tg$ defined by an integral function of the Brunt-V\"ais\"al\"a
frequency in the inner radiative region. The measured values of
$\Tg$ are accurately reproduced for red giant ascending the RGB,
but not in the clump \citep{2013ApJ...766..118M}. This discrepancy
in the clump is not yet understood and deserves further work.

\begin{figure}[t!]
 \centering
 \includegraphics[width=0.8\textwidth,clip]{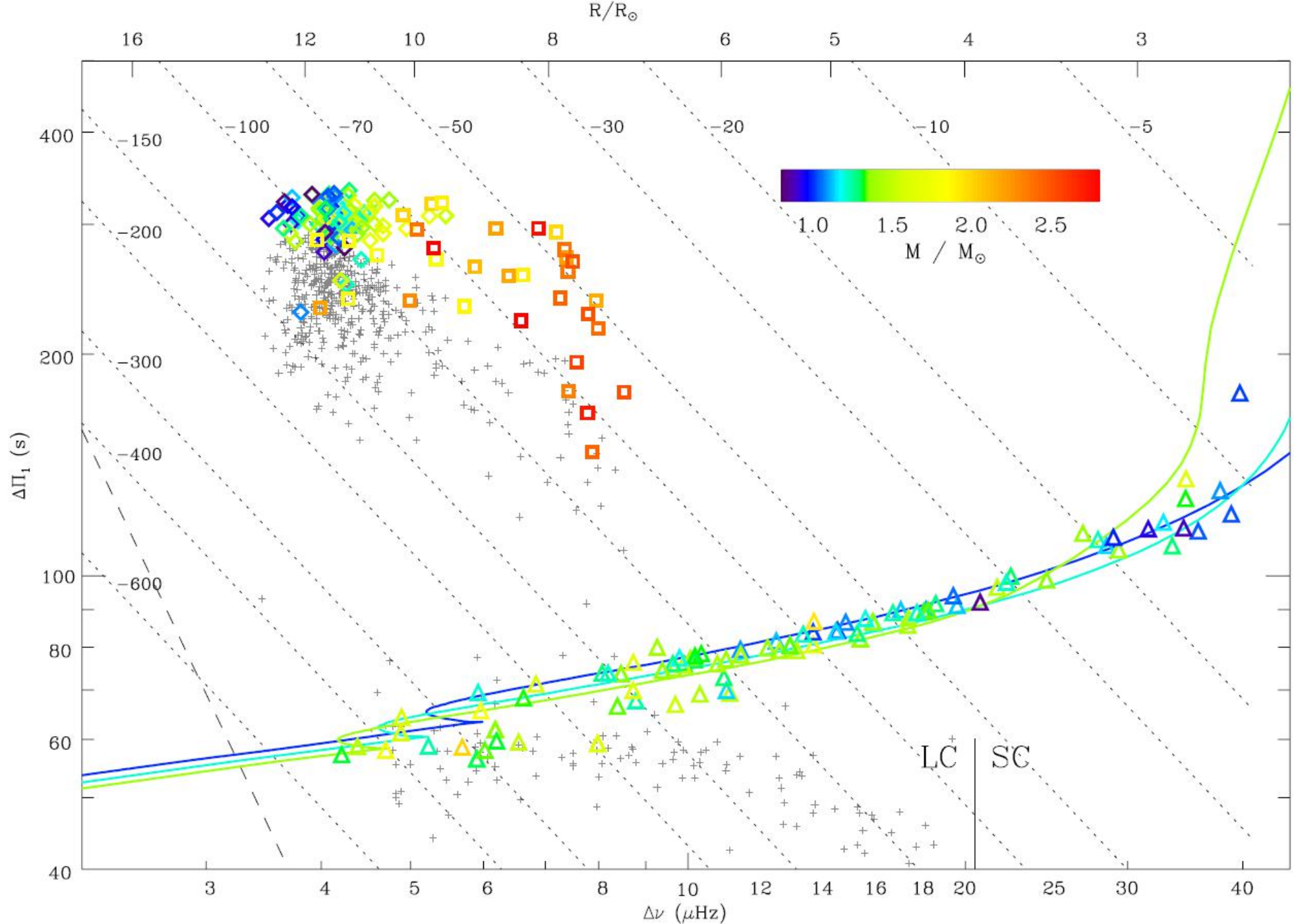}
  \caption{Gravity-mode period spacing $\Tg$ as a function of the
pressure-mode large frequency spacing $\Dnu$, for a set of red
giants observed with \Kepler. Long-cadence data (LC) have $\Dnu\le
20.4\,\mu$Hz. RGB stars are indicated by triangles; clump stars by
diamonds; secondary clump stars by squares. Uncertainties in both
parameters are smaller than the symbol size. The seismic estimate
of the mass is given by the color code. Small gray crosses
indicate the bumped periods $\Delta P$ measured by
\cite{2011A&A...532A..86M}. Dotted lines are $\ng$ isolines. The
dashed line in the lower left corner indicates the formal
frequency resolution limit. The upper x-axis gives an estimate of
the stellar radius for a star whose $\numax$ is related to $\Dnu$
according to the mean scaling relation $\numax =
(\Dnu/0.28)^{1.33}$ (both frequencies in $\mu$Hz). The solid
colored lines correspond to a grid of stellar models with masses
of 1, 1.2 and $1.4\, M_\odot$, from the ZAMS to the tip of the
RGB.   Figure from \cite{2012A&A...540A.143M}.}
  \label{DnuDPi1}
\end{figure}

Homology seen in the radial red giant oscillation pattern is also
seen in the $\Dnu$ -- $\Tg$ diagram (Fig.~\ref{DnuDPi1}). All
low-mass stars on the RGB lie on the same track, independent of
their mass and composition. With $\Dnu$ representing the stellar
mean density and $\Tg$ representing the density of the core, we
derive from Fig.~\ref{DnuDPi1} that the properties of the core and
of the envelope are closely correlated. The inert helium core of
an RGB star necessarily contracts when the star ascends the RGB;
accordingly, its mass increases, as a result of the hydrogen
burning shell which surrounds it. Its contraction and mass
increase yield a decreasing $\Tg$ term, as observed and as seen in
the modeling \citep{2013ApJ...766..118M}. This contraction implies
that the density of the hydrogen burning shell increases. This
increase boosts the energy production. Hence, the hydrogen
envelope expands.

Homology is reinforced in the red clump, since a further similar
event has participated to erase the initial differences between
low-mass stars. Their helium core being degenerate, the helium
flash occurs in very similar conditions
\citep{2013ApJ...766..118M}, so that they nearly reach the same
location in the $\Tg$ -- $\Dnuobs$ diagram, close to $\Tg \simeq
300\,$s and $\Dnu\simeq 4\,\mu$Hz.

The precision we have on the mixed mode determination is so high
that we may expect, from an accurate modelling, a highly precise
age determination. During the evolution on the RGB, the gravity
period spacing $\Tg$ change is about 100\,s (Fig. \ref{DnuDPi1}).
Since the measurement of $\Tg$ is more precise than 0.1\,s, it is
formally possible to track the timing of the ascent of a red giant
with a precision as high as 0.1\,\%.

\begin{table}
\caption{Scaling relations}\label{table-fit}
\begin{tabular}{lllccl }
\hline parameter &  & unit & coefficient $\alpha$ & exponent $\beta$\\
\hline
large separation & $\Dnuobs$  & $\mu$Hz           &$0.274\pm 0.004$ & $0.757\pm0.004$ & \\
                 &$\nmax=\numax/\Dnu -\epsobs$ &--&$3.26\pm0.031$   & $0.242\pm0.005$ & \\
FWHM             & $\dnuenv$  & $\mu$Hz           &$0.73 \pm 0.03 $ & $0.88\pm0.01$   & \\
                 & $\nenv= \dnuenv/\Dnu$       &--&$2.49 \pm 0.12$  & $0.13\pm0.01$ & \\
Height at $\numax$    &$\Hmax$&ppm$^2$\,$\mu$Hz$^{-1}$&$(2.03\pm 0.05)\;10^7$& $-2.38\pm0.01$ &  \\
Background at $\numax$&$\Bmax$&ppm$^2$\,$\mu$Hz$^{-1}$&$(6.37\pm 0.02)\;10^6$& $-2.41\pm0.01$ &  \\
HBR              &$\Hmax / \Bmax$              &--&$3.18\pm 0.09$   & $0.03\pm 0.03$  \\
Granulation      & $P\ind{g}$&ppm$^2$\,$\mu$Hz$^{-1}$ &                      & $-2.15\pm0.12$ &  \\
                 &$\tau\ind{g}$& s &                   & $-0.90\pm0.005$&\\
  &$P\ind{g}(\tau\ind{g})$&ppm$^2$\,$\mu$Hz$^{-1}$&        & $2.34\pm0.01$ &  \\
 \hline
radius & $R$        &   $R_\odot$                  &  $63.1\pm 1.1$  & $-0.49\pm 0.01$ \\
effective temperature& $\Teff$    & K              &  $3922\pm 50$   & $0.051\pm 0.05$ \\
\hline
\end{tabular}

\scriptsize{
- All results were obtained with the COR pipeline \citep{2012A&A...537A..30M}. Granulation data are from
\cite{2011ApJ...741..119M}.

- Each parameter is estimated as a power law of
$\numax$, with $\alpha$
the coefficient and $\beta$ the exponent, except $P\ind{g}(\tau\ind{g})$.

}
\end{table}

\section{Scaling relations\label{scaling}}

\subsection{Seismic parameters}

As a result of homology, the red giant global seismic  parameters
conform to a large numbers of scaling relations.  Their variations
with $\numax$  are summarized in Table~\ref{table-fit}, where we
consider:

- The mean observed large separation $\Dnuobs$ measured in a broad
frequency range around $\numax$, and $\nmax = \numax / \Dnuobs -
\epsobs$ which provides an estimate of the radial order at
$\numax$; $\nmax$  significantly decreases when $\Dnuobs$
decreases. For the Sun, $\nmax\simeq 22$; at the red clump,
$\nmax\simeq 8$; and at the tip of the RGB, $\nmax\simeq 2.5$.

- $\dnuenv$ is the full-width at half-maximum of the smoothed
excess power; $\nenv = \dnuenv / \Dnuobs$ provides $\dnuenv$ is
large separation unit; as $\nmax$, $\nenv$ significantly decreases
when $\Dnuobs$ decreases since $\dnuenv$ approximately scales as
$\numax$.

- $\Hmax$ (in ppm$^2\,\mu$Hz$^{-1}$) is the mean height of the
modes at $\numax$, defined according to the description of
smoothed excess power as a Gaussian envelope
\citep[e.g.,][]{2012A&A...537A..30M}.

- $\Bmax$ (in ppm$^2\,\mu$Hz$^{-1}$) is the value of the stellar
background $B$ at $\numax$. The background is described by
Harvey-like components \citep{2008Sci...322..558M}. Each component
is a modified Lorentzian of the form $b(\nu) = a / [1 +
(2\pi\,\nu\tau)^{\alpha}]$, where $\tau$ is the characteristic
time scale. The exponent $\alpha$ is in the range 2 -- 4
\citep{2011ApJ...741..119M}. $\Hmax / \Bmax$ is representative of
the height to background ratio (HBR) at $\numax$. This ratio shows
no significant variation all along the RGB.

- Properties of the granulation signal are also considered
\citep{2011ApJ...741..119M}: $P\ind{g}$ is the height of the
granulation component ; $\tau\ind{g}$ is the time scale of this
background component related to the granulation signal. We note
that the exponent of the $P\ind{g}(\tau\ind{g})$ scaling relation
is, in absolute value, very close to the exponent of the $\Bmax
(\numax)$ scaling relation. This means that the energy
content in the granulation and in the oscillations are certainly
linked.

- Finally, we provide estimates of the fundamental parameters $R$
and $\Teff$.

\begin{figure}[t!]
 \centering
 \includegraphics[width=0.999\textwidth,clip]{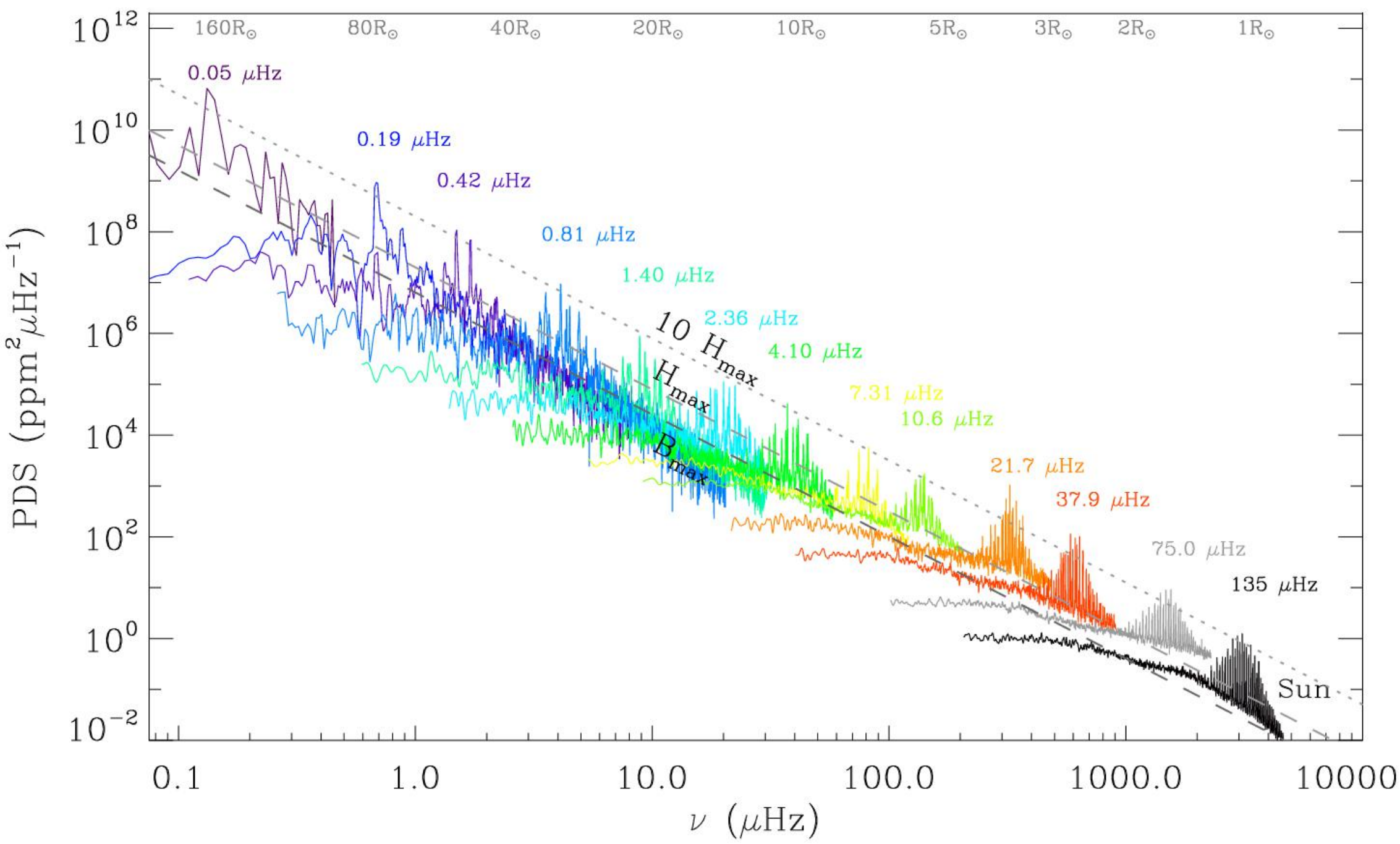}
\caption{Superimposition of red giant oscillation spectra from the
bottom to the tip of the RGB. Each spectrum is identified by the
large separation $\Dnuobs$. The Sun spectrum (in black) and
another main-sequence spectrum (in grey) are shown for comparison.
Except at very low frequency, all spectra were filtered, with a
filter width varying with $\numax$. The fit of the background and
mean height at $\numax$ (Table~\ref{table-fit}) are indicated with
dashed lines. A proxy of the stellar radius is provided along the
upper frequency ($\numax$) axis. Data from
\cite{2012A&A...540A.143M} and \cite{mosser_ogle}}
  \label{fig-scaling}
\end{figure}

Some of the scaling relations are illustrated in
Fig.~\ref{fig-scaling} that shows, compared to the Sun,
oscillation spectra of red giants from the bottom to the top of
the RGB. Currently, we lack theoretical models for explaining most
of these relations. Large efforts have been devoted to explain the
scaling relations  of the mean amplitude $\Amax$.  This global
parameter can be fitted, in limited frequency range, as in
\cite{2012A&A...537A..30M}. However, the fit heavily depends on
the method \citep{2011ApJ...737L..10S,2011ApJ...743..143H}, so
that it is not yet possible to provide a physically relevant
result \citep{2013MNRAS.430.2313C}. \cite{2012A&A...543A.120S}
have shown that scaling relations of mode amplitudes cannot be
extended from main-sequence to red giant stars because
non-adiabatic effects for red giant stars cannot be neglected.
\cite{2013arXiv1309.1620S} have recently proposed a theoretical
model of the oscillation spectrum associated with the stellar
granulation as seen in disk-integrated intensity. With this model,
\cite{2013arXiv1309.1488S} have highlighted the role of the
photospheric Mach number for controlling the properties of the
stellar granulation.

\subsection{Mass and radius scaling relations}

As already stated, the seismic global parameters $\numax$ and
$\Dnu$ scale, respectively, as the atmospheric cutoff frequency
and as the square root of the mean density. Hence, seismic
relations with $\Dnuobs$ and $\numax$ can be used to provides
proxies of the stellar masses and radii. Following
\cite{2013A&A...550A.126M}, we stress that it is necessary to
avoid confusion between the large separation $\Dnuobs$ observed
around $\numax$, the asymptotic large separation $\Dnuas$, and the
dynamical frequency $\nu_0$ that scales with
$\sqrt{\mathcal{G}M/R^3}$. The measurement of $\Dnuobs$ is
perturbed by frequency glitches due to the rapid local variation
of the sound speed in the stellar interior, related to the density
contrast at the core boundary or to the local depression of the
sound speed that occurs in the helium second-ionization region
\citep{2010A&A...520L...6M}. The scaling relations write then:
\begin{equation}
  {\Rsis \over\Rs}  = \left({\numax \over \numaxas}\right) \
     \left({\Dnuas \over \dnusas}\right)^{-2}
     \left({\Teff \over \Ts}\right)^{1/2}, \label{scalingRas}
\end{equation}
\begin{equation}
  {\Msis\over\Ms} = \left({\numax \over \numaxas}\right)^{3}
     \left({\Dnuas \over \dnusas}\right)^{-4} \left({\Teff \over \Ts}\right)^{3/2} , \label{scalingMas}
\end{equation}
with the calibrated references $\numaxas = 3104\,\mu$Hz and
$\dnusas = 138.8\,\mu$Hz. Such reference values were determined in
order to avoid systematic bias between the seismic proxies and modeled values \citep{2013A&A...550A.126M}.

Compared to scaling relations heavily used elsewhere
\citep[e.g.,][]{2010A&A...517A..22M,2010A&A...522A...1K,2011ApJ...732...54C,2011ApJ...738L..28V,2011ApJ...740L...2S,2013ApJ...765L..41S,2013A&A...556A..59H},
Eqs.~\ref{scalingRas} and \ref{scalingMas} make use of the
asymptotic large separation instead of the observed large
separation. They introduce a correction which can be expressed by
\begin{equation}\label{corR}
  \Ras \simeq  \bigl[{ 1 - 2\, (\zeta-\zeta_\odot)} \bigr] \Robs
  \hbox{ \ and \ }
  \Mas \simeq  \bigl[{ 1 - 4\, (\zeta-\zeta_\odot)} \bigr]
  \Mobs ,
\end{equation}
with $\zeta=0.57/\nmax$ in the main-sequence regime and $\zeta =
0.038$ in the red giant regime. The amplitude of the correction
takes into account the fact that scaling relations are calibrated
on the Sun, so that one has to deduce the solar correction
$\zeta_\odot \simeq 0.026\,$\%. These asymptotic corrections
represent a first step towards the proper calibration of the
relation. Contrary to the forms based on the observed large
separation, they are not biased, and free of the perturbation of
the glitches \citep{roscoff}. This result might contradict
\cite{2013arXiv1307.3132B}, who state that $\Dnuobs$ provides a better proxy
of $\nu_0$ than $\Dnuas$. In fact, the contradiction is apparent
only: the calibration effort has still to link the asymptotic
values derived from the models or derived from the observations
(Table~\ref{table-calibration}).

\begin{table}[t]
\centering
\caption{Calibration of the mass and radius scaling relations}\label{table-calibration}
\begin{tabular}{rrcl}
\hline
 & Modeling &   Calibration & Observation\\
\hline
$M,R\leftrightarrow$ & $\Dnuasmod \leftrightarrow$ & ? & $\leftrightarrow \Dnuasobs \leftrightarrow \Dnuobs$\\
                     & $\numaxmod \leftrightarrow$ & ? & $\leftrightarrow \numaxobs $\\
\hline
\end{tabular}

\scriptsize{Calibration process unveiling all formal steps for a
proper calibration of the mass and radius scaling relations, with
checking of the asymptotic values, determined in the modeling
process from $1/(2\int_0^R \diff r / c)$ and in observations from
the asymptotically corrected  $\Dnuobs$.
}
\end{table}

\subsection{Calibration of the mass and radius scaling relations}

For a proper calibration of the mass and radius scaling relations,
we should use the dynamical frequency $\nu_0$ instead of $\Dnuobs$
or $\Dnuas$, and the acoustic frequency $\nu\ind{c}$ instead of
$\numax$.  As this is not the case, an intensive calibration
effort is undertaken:

- An independent verification has been made for stars that have
accurate Hipparcos parallaxes, by  coupling asteroseismic analysis
with the InfraRed Flux Method \citep{2012ApJ...757...99S}. The
seismic distance determinations agree to better than 5\,\%: this
shows the relevance and the accuracy of the scaling relations in
the subgiant and main-sequence regime.

- With long-baseline interferometric measurement of the radius of
five main-sequence stars, one subgiant and four red giant stars
for which solar-like oscillations have been detected by either
Kepler or CoRoT, \cite{2012ApJ...760...32H} have shown that
scaling relations are in excellent agreement within the
observational uncertainties. They finally derive that
asteroseismic radii for main-sequence stars are accurate to better
than 4\,\%.

- Oscillations in cluster stars \citep{2011ApJ...729L..10B} were
used to compare scaling relations for red giants  in the red clump
or on the RGB. \cite{2012MNRAS.419.2077M} have found evidence for
systematic differences in the $\Dnuobs$ scaling relation between
He-burning and H-shell-burning giants. This implies that a
relative correction between RGB and clump stars must be
considered. As this correction is also related to mass loss, it is
currently not possible to measure it precisely. Independent of
this, oscillations in cluster stars provide useful constraints on
membership \citep{2011ApJ...739...13S}. They also help constrain
the relations depicting the parameters of the pressure mode
spectrum \citep{2012ApJ...757..190C}.

\begin{figure}[t!]
 \centering
 \includegraphics[width=0.65\textwidth,clip]{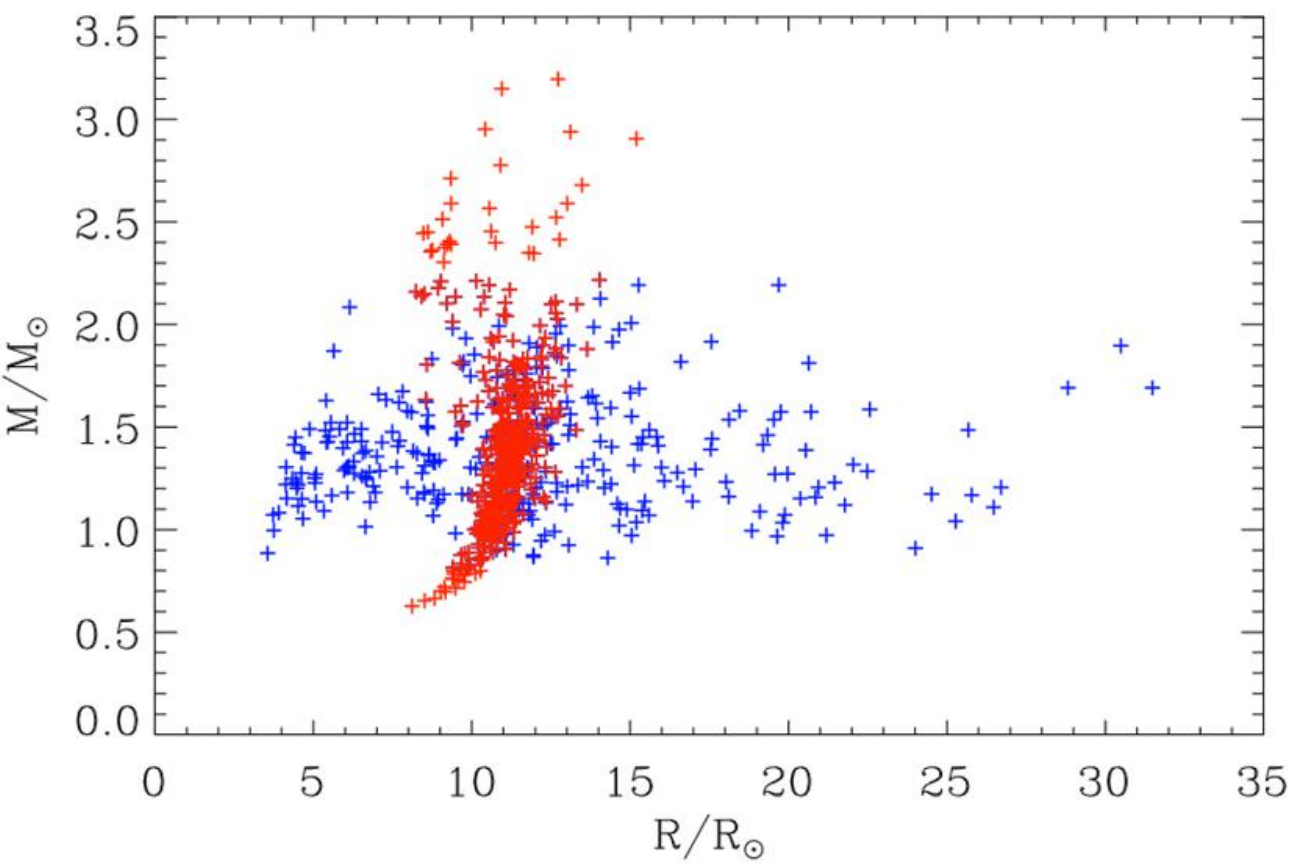}%
  \caption{Mass -- radius relation for \Kepler\ red giants, with RGB stars in blue and clump stars in red.
  Adapted from \cite{2012A&A...537A..30M}.}
  \label{figMR}
\end{figure}

\subsection{Radius, mass and mass loss\label{MR}}

The mass and radius scaling relations are illustrated by a mass --
radius diagram (Fig.~\ref{figMR}). The uncertainties on the
inferred mass and radius values discussed above warn us that the
absolute values of the data reported here are not yet fully
calibrated. However, the trends and the relative variations are
relevant, so that it is possible to derive unique information:

- Most of the stellar masses observed on the RGB are in the range
[1, $2\,M_\odot$]; less-massive stars spend a longer lifetime on
the main sequence, so that they need time to reach the RGB; more
massive stars are intrinsically rare and evolve quickly on
different evolutionary tracks, especially in the instability
strip. So, many of them do not experience solar-like oscillations
before having met the RGB; they reach it with a larger radius than
lower-mass stars;

- In the red clump, consequently after the tip of the RGB where
strong episodes of mass loss occur, the mass distribution is
significantly different. Stars with mass down to $0.6\,M_\odot$
are observed: they have necessarily suffered from efficient mass
loss near the tip of the RGB. Stars with mass over $2\,M_\odot$
are present, since they spend more time in the helium-burning
phase than ascending the RGB in the region where solar-like
oscillations are observed.

- The mass-radius relation in the red clump shows a limited
spread. All stars in this stage share common properties, as
derived from the examination of the $\Tg$ -- $\Dnuobs$ relation on
the RGB (Fig.~\ref{DnuDPi1} and Section \ref{mixtes}). If we add
the information of the effective temperature in Fig.~\ref{figMR},
we note that, at fixed mass, the hottest stars have the smallest
radius, in agreement with a thinner envelope.

- Secondary-clump stars show a larger spread in the mass-radius
diagram compared to clump stars. The difference may arise from the
ignition of helium having started in non-degenerate conditions. As
already stated above, the mass of transition between the primary
and secondary clumps, about $1.8\,M_\odot$, is indicative only.
Its precise determination requires the careful calibration of the
scaling relation (Eq.~\ref{scalingMas}).

\begin{figure}[t!]
 \centering
 \includegraphics[width=0.8\textwidth,clip]{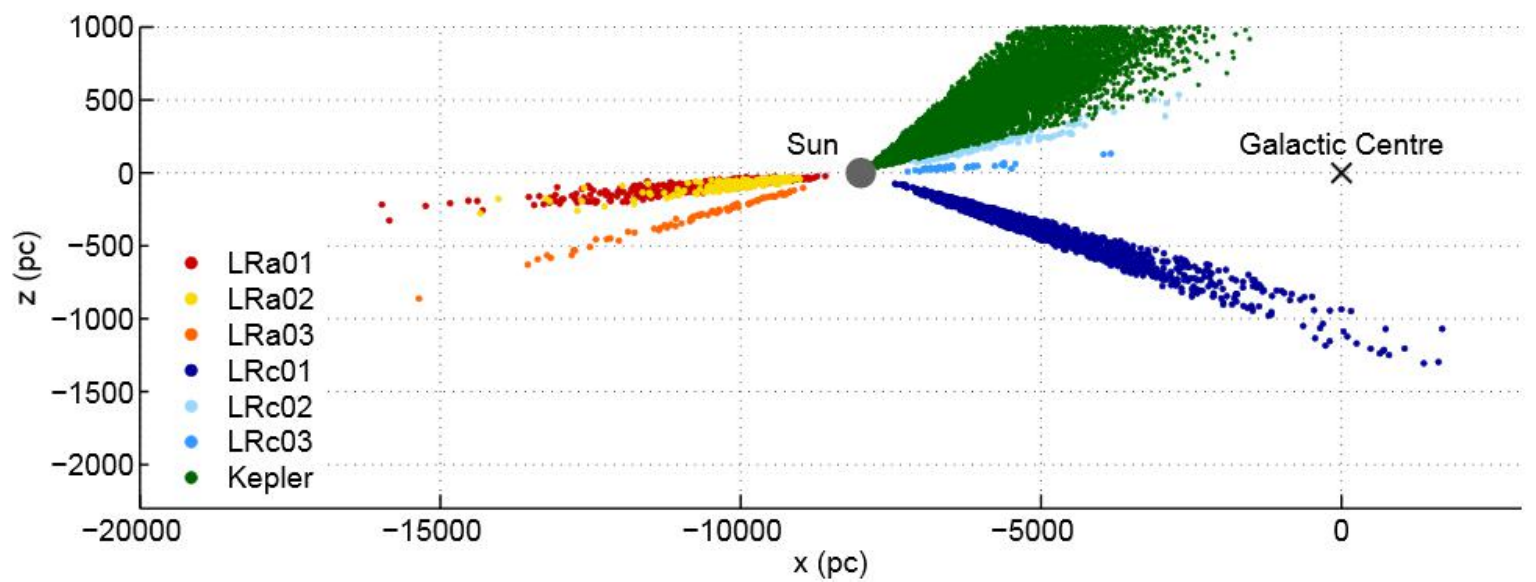}
\caption{Solar-like oscillating red giants observed in several
CoRoT fields of view and by \Kepler : projection on the $x$ -- $z$
plane. Figure from \cite{2012EPJWC..1905012M}. \label{fig-popu}}
\end{figure}

\subsection{Distance measurements and stellar population\label{distancepop}}

An important consequence of the measurement of asteroseismic radii
for field stars is the capability of measuring stellar distance.
\cite{2013MNRAS.429..423M} have shown that red giants represent a
well-populated class of accurate distance indicators, spanning a
large age range, which can be used to map and date the Galactic
disk in the regions probed by observations made by the CoRoT and
\Kepler. They have determined precise distances for ˜2000 stars
spread across nearly 15\,000\,pc of the Galactic disk, exploring
regions which are a long way from the solar neighbourhood
(Fig.~\ref{fig-popu}). Significant differences in the mass
distributions of these two samples are interpreted as mainly due
to the vertical gradient in the distribution of stellar masses
(hence ages) in the disk.

\begin{figure}[t!]
 \centering
 \includegraphics[width=0.8\textwidth,clip]{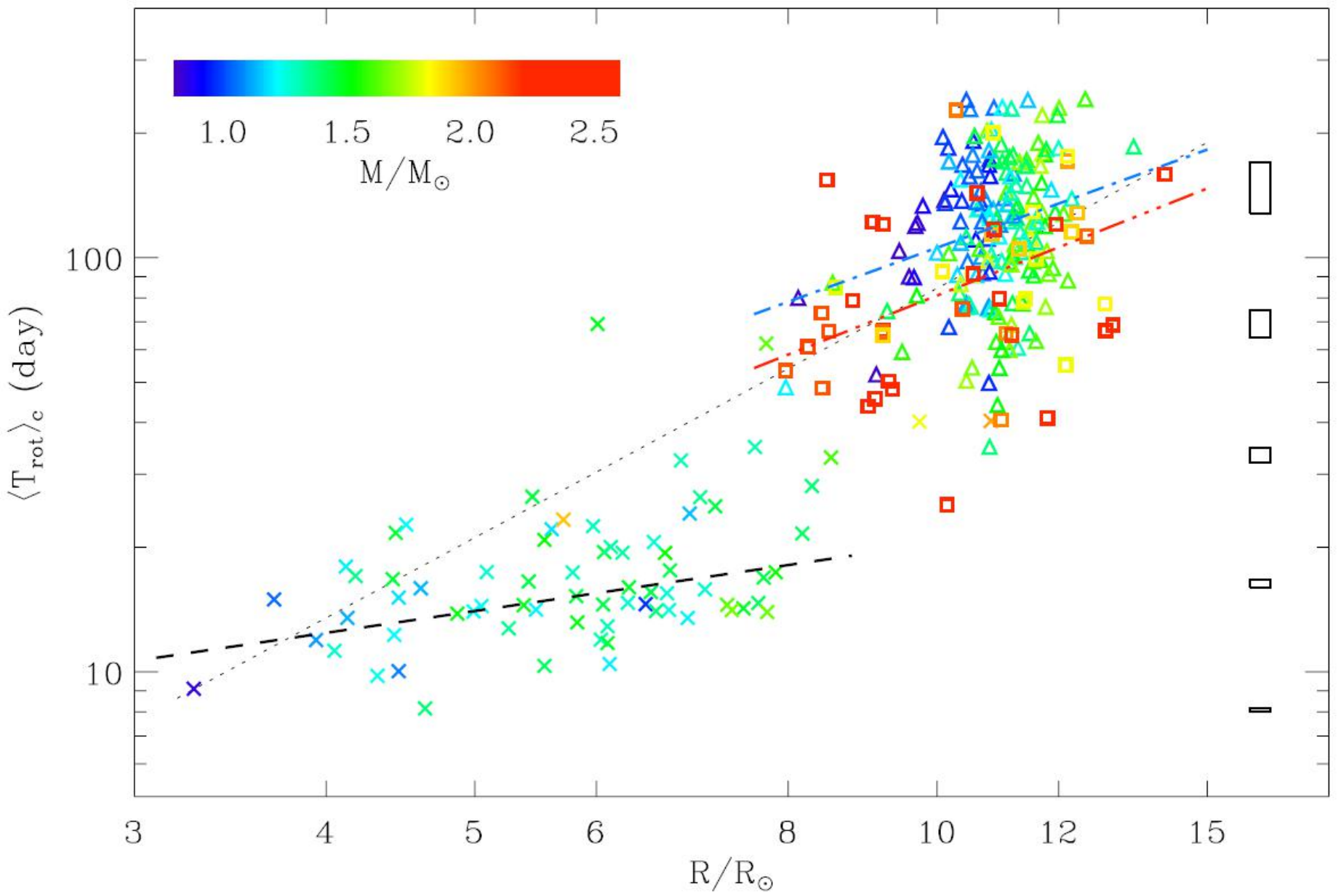}
  \caption{Mean period of core rotation as a function of the
asteroseismic stellar radius, in log-log scale. The dotted line
indicates a rotation period varying as $R^2$. The dashed
(dot-dashed, triple-dot-dashed) line indicates the fit of RGB
(clump, secondary clump) core rotation period. The rectangles in
the right side indicate the typical error boxes, as a function of
the rotation period. Figure from \cite{2012A&A...548A..10M}.
\label{mean-rotation}}
\end{figure}

\section{Rotation\label{rotation}}

Rotational splittings have been first observed in a handful of red
giants, putting in evidence a significant radial differential
rotation \citep{2012Natur.481...55B,2012ApJ...756...19D}.
\cite{2012A&A...540A.143M} have developed a dedicated method for
automated measurements of the rotational splittings in a large
number of red giants. They have also shown that these splittings,
dominated by the core rotation, can be modeled with a Lorentzian
function that resembles the mixed-mode pattern organization. Under
the assumption that a linear analysis can provide the rotational
splitting, they note a small decrease of the mean core rotation
rate of stars ascending the RGB. Alternatively, an important slow
down is observed for red-clump stars compared to the RGB. They
also show that, at fixed stellar radius, the specific angular
momentum increases with increasing stellar mass
(Fig.~\ref{mean-rotation}).

With the same hypothesis of linear splittings,
\cite{2013A&A...549A..75G} have described the morphology of the
rotational splittings. They have proven that the mean core
rotation dominates the splittings, even for pressure dominated
mixed modes. For red giant stars with slowly rotating cores, the
variation in the rotational splittings of dipole modes with
frequency depends only on the large frequency separation, the
g-mode period spacing, and the ratio of the average envelope to
core rotation rates. Thus, they have proposed a method to infer
directly this ratio from the observations and have validated this
method using \Kepler\ data. In case of rapid rotation, rotation
cannot be considered as a perturbation any more and the linear
approach fails \citep{2013A&A...554A..80O}.

For investigating the internal transport and surface loss of the
angular momentum of oscillating solar-like stars,
\cite{2013A&A...549A..74M} have studied the evolution of
rotational splittings from the pre-main sequence to the red-giant
branch (RGB) for stochastically excited oscillation modes. They
have shown that transport by meridional circulation and shear
turbulence cannot explain the observed spin-down of the mean core
rotation. They suspect the horizontal turbulent viscosity  to be
largely underestimated.

\begin{figure}[t!]
 \centering
 \includegraphics[width=0.8\textwidth,clip]{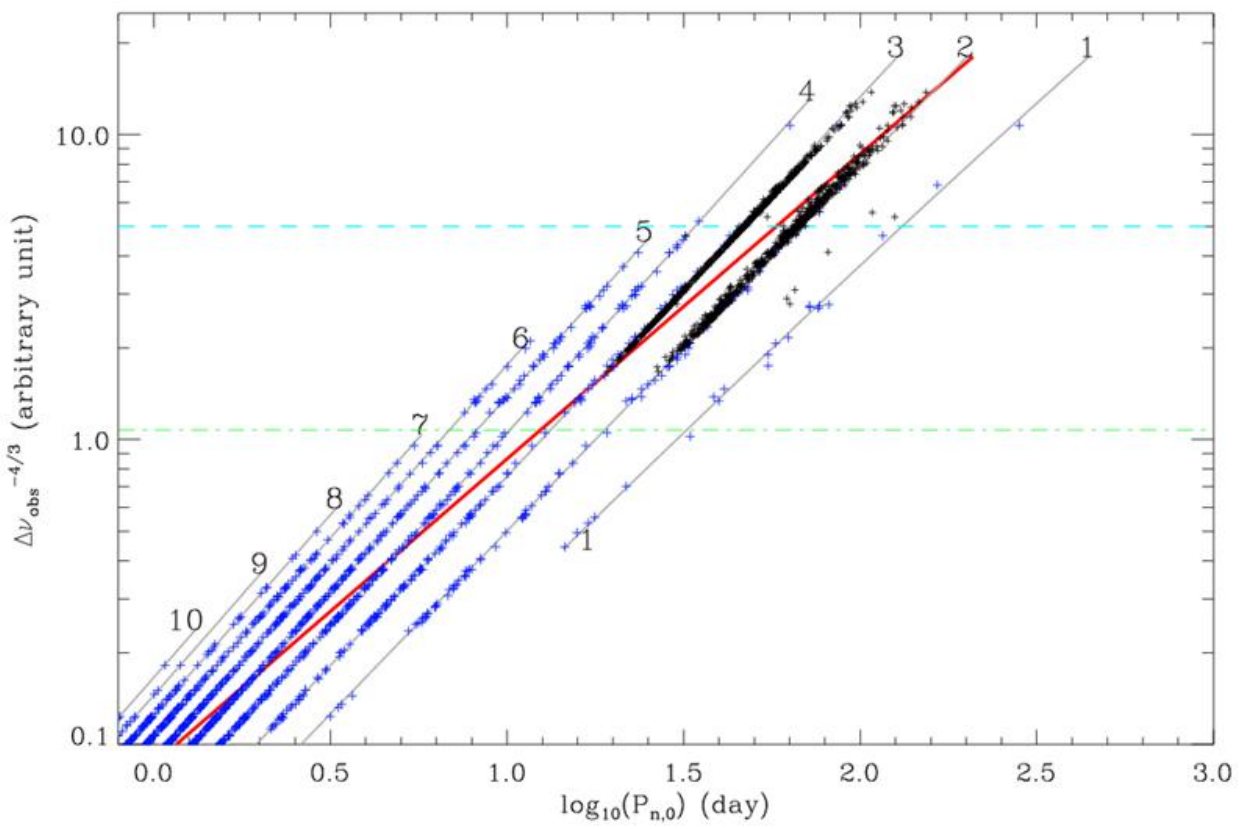}
  \caption{Period-luminosity relations with \Kepler \ (blue) and OGLE data (black),
  from \cite{mosser_ogle}. The blue dashed line indicates the location of the tip of the RGB.}
  \label{fig-ogle}
\end{figure}

\section{Oscillations in evolved M giants\label{ogle}}

Semi-variability in evolved M giants is suspected to be due to
solar-like oscillations. However, until recently, only indirect
information was available for sustaining this hypothesis
\citep[e.g.,][]{2010A&A...524A..88D}. The question concerning the
nature of these oscillations is now solved with \Kepler\
observations \citep{mosser_ogle}. According to scaling relations,
such oscillations occur at very low frequency: at the tip of the
RGB, $1/\numax$ corresponds to periods of 50 days.  The monitoring
of such oscillations has benefitted from the unique length of
\Kepler\ observation (more than 3 years) and is unfortunately out
of reach with CoRoT observations limited to five months
\citep[e.g.,][]{2009A&A...506...51B}.

The solution for unambiguously identifying solar-like oscillations
at very low frequency is based on two arguments. First, the
relevance of the universal red giant oscillation pattern for
less-evolved evolutionary stages implies that, at late stages,
oscillation spectra very certainly show homologous properties.
Second, the $\epsobs(\Dnuobs)$ relation observed in the whole red
giant oscillation regime is justified by the validity of the
second-order asymptotic expansion. Consequently, this relation was
extrapolated to very low $\numax$, and iteratively adapted to
provide an acceptable fit of the M-giant oscillation spectra. The
success of the fits for all red giants, except in a limited number
of cases with a very low signal-to-noise-ratio oscillation
spectrum or frequency leakage due to binarity, has proven the
relevance of the method.

Period-luminosity relations interpreted as solar-like oscillations
are shown in Fig.~\ref{fig-ogle}; OGLE data are superimposed. When
the large separation decreases, the radial orders of the observed
modes decrease too, down to $\nmax=2$.  The fits show that mostly
radial $n=2$ and 3 modes are observed below the tip of the RGB.
Interpreting oscillations in semi-regular variables has many
consequences: the parametrization of the oscillation spectrum will
help reanalyze the ground-based observations and to define more
accurately the different sequences. Interpreting period-luminosity
relations in red giants in terms of solar-like oscillations might
by used to reinvestigate with a firm physical basis the time
series obtained from ground-based microlensing surveys. This will
provide improved distance measurements and open the way to
extragalactic asteroseismology, with the observations of M giants
in the Magellanic Clouds. \cite{mosser_ogle} have also shown that
the acceleration of the external layers of red giant with
solar-like oscillations is about the same order of magnitude as
the surface gravity when the stars reach the tip of the RGB. This
shows that oscillations might play a non-negligible role in the
mass-loss process.


\section{Conclusions?}

Ensemble asteroseismology is an active field in current progress.
Any conclusion written now will be out of date tomorrow. Plenty of
work remains to be done:

- examining the fine structure of the universal red-giant
oscillation pattern;

- establishing a thorough calibration of the mass and radius
scaling relations;

- modelling a large number of stars, with improved stellar
physics;

- deriving precise stellar ages;

- irrigating many connected themes: distance measurement,
population study, gyrochronology, late stages evolution...

Following the successful space missions CoRoT and \Kepler, a
next-generation seismic project requires simple but demanding
characteristics: long, continuous, and ultra-precise photometric
observations. A complete survey of the sky, twin of the ESA Gaia
mission, able to derive the seismic global indices $\Dnuobs$ and
$\numax$ for millions of stars, is certainly a highly-promising
project.

\begin{acknowledgements}
Funding for the Discovery mission \Kepler\ is provided by NASA's
Science Mission Directorate. The authors acknowledge financial
support from the ``Programme National de Physique Stellaire"
(PNPS, INSU, France) of CNRS/INSU and from the ANR program IDEE
``Interaction Des \'Etoiles et des Exoplan\`etes'' (Agence
Nationale de la Recherche, France).
\end{acknowledgements}

\renewcommand{\baselinestretch}{.9}
\bibliographystyle{aa}  
\bibliography{mosser_sf2a} 

\end{document}